\documentclass[twocolumn,english]{revtex4}
\usepackage[T1]{fontenc}
\usepackage[latin9]{inputenc}
\usepackage{textcomp}
\usepackage{amstext}
\usepackage{graphicx}
\usepackage{esint}

\makeatletter

\newcommand{\lyxmathsym}[1]{\ifmmode\begingroup\def\b@ld{bold}
  \text{\ifx\math@version\b@ld\bfseries\fi#1}\endgroup\else#1\fi}

\@ifundefined{textcolor}{}
{%
 \definecolor{BLACK}{gray}{0}
 \definecolor{WHITE}{gray}{1}
 \definecolor{RED}{rgb}{1,0,0}
 \definecolor{GREEN}{rgb}{0,1,0}
 \definecolor{BLUE}{rgb}{0,0,1}
 \definecolor{CYAN}{cmyk}{1,0,0,0}
 \definecolor{MAGENTA}{cmyk}{0,1,0,0}
 \definecolor{YELLOW}{cmyk}{0,0,1,0}
}

\makeatother

\usepackage{babel}
\begin{document}

\title{Numerical calculation of magnetic form factors of complex shape nano-particles
coupled with micromagnetic simulations. }

\author{Fatih Zighem$^{1,2}$, Frédéric Ott$^{2,3}$}

\email{Email: frederic.ott@cea.fr, }

\selectlanguage{english}%

\author{Thomas Maurer$^{3,4}$, Grégory Chaboussant$^{2,3}$, Jean-Yves Piquemal$^{5}$
and Guillaume Viau$^{6}$}

\affiliation{$^{1}$LSPM, CNRS-Université Paris XIII, 93430 Villetaneuse, France}

\affiliation{$^{2}$CEA, IRAMIS, Laboratoire Léon Brillouin, 91191 Gif sur Yvette,
France }

\affiliation{$^{3}$CNRS, IRAMIS, Laboratoire Léon Brillouin, 91191 Gif sur Yvette,
France}

\affiliation{$^{4}$LNIO, CNRS-Université de technologie de Troyes, 10000 Troyes,
France}

\affiliation{$^{5}$ITODYS, CNRS-Université Paris VII,75205 Paris, France}

\affiliation{$^{6}$LPCNO, CNRS-INSA-Université de Toulouse, 31077 Toulouse, France}
\begin{abstract}
We investigate the calculation of the magnetic form factors of nano-objects
with complex geometrical shapes and non homogeneous magnetization
distributions. We describe a numerical procedure which allows to calculate
the 3D magnetic form factor of nano-objects from realistic magnetization
distributions obtained by micromagnetic calculations. This is illustrated
in the canonical cases of spheres, rods and platelets. This work is
a first step towards a 3D vectorial reconstruction of the magnetization
at the nanometric scale using neutron scattering techniques. 
\end{abstract}


\maketitle

\section{Introduction }

\begin{figure*}
\includegraphics[bb=70bp 420bp 760bp 570bp,clip,width=15cm]{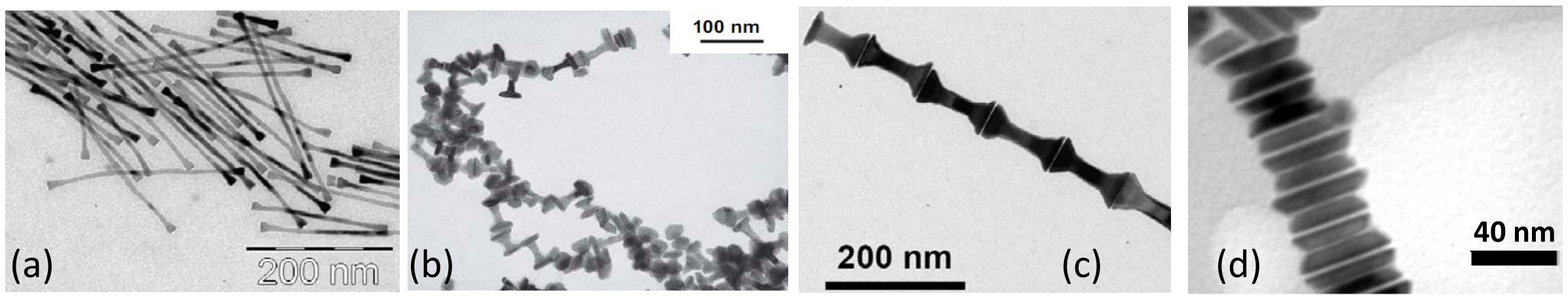}
\caption{TEM images of magnetic nano-objects synthesized in various conditions:
(a) $Co_{80}Ni_{20}$ nanowires with rounded tips of length $200$
nm and diameter $7$ nm ; (d) $Co_{50}Ni_{50}$ dumbbells ; (e) $Co$
diabolos and (f) \emph{$Co$} platelets.}

\label{Fig1:images} 
\end{figure*}

The recent progress in solid state chemistry has led to the possibility
of synthesizing nano-objects of non spherical shapes: wires by organometallic
chemistry \cite{soulantica2009} or electrochemistry \cite{wade2005}.
In particular, the use of the polyol process has made it possible
to produce well defined monodisperse magnetic nano-objects. Depending
on the synthesis conditions, various shapes of particles can be obtained
such as rods, wires, dumbbells, diabolos, platelets... (see Figure
\ref{Fig1:images}) \cite{viau2008,soumare2008,soumare2009}. These
are the typical forms of magnetic nano-objects which will be the focus
of this communication. While magnetic nanospheres (as found for example
in ferrofluids) have been extensively studied by Small Angle Neutron
Scattering \cite{michels_rev,Wiedenmann2003,Wiedenmann2004,wiedenmann2006,michels2005,michels2006},
more complex shaped magnetic nano-objects have rarely been studied.

In this communication we focus on the detailed description and calculation
of neutron scattering on magnetic nano-objects in SANS experiments.
We describe numerical tools to calculate the magnetic form factors
of arbitrary shape nano-objects. We present a practical procedure
which allows to calculate magnetic form factors either from an a priori
knowledge of the magnetization distribution or from a minimization
of the micro-magnetic configuration.

\section{Nuclear and magnetic form factors calculations for neutrons}

In this section we first recall the interaction of neutrons with small
nano-objects. We introduce the nuclear and magnetic form factors of
nano-objects and describe the quantities which can be measured in
a Polarized Small Angle Neutron Scattering (PSANS) experiment. A practical
way of calculating the magnetic form factor of complex nano-objects
is then presented.

\subsection{Nuclear and magnetic form factors}

Let us consider a magnetic nano-object of nuclear Scattering Length
Density (SLD) $\rho\left(\vec{r}\right)$ which creates an induction
distribution $\overrightarrow{B}\left(\vec{r}\right)$. We define
the nuclear form factor $f_{N}(\vec{Q})$ as the Fourier transform
of the nuclear SLD distribution, $\vec{Q}=\vec{k_{f}}-\vec{k_{i}}$
being the scattering wave-vector:

\begin{equation}
f_{N}\left(\vec{Q}\right)\propto\iiint\rho\left(\vec{r}\right)e^{-i\vec{Q}.\vec{r}}d\vec{r}
\end{equation}

Similarly we define the magnetic form factor $\vec{f}_{B}\left(\vec{Q}\right)$
as the Fourier transform of the induction field $\overrightarrow{B}\left(\vec{r}\right)$:

\begin{equation}
\vec{f}_{B}\left(\vec{Q}\right)\propto\iiint\vec{B}\left(\vec{r}\right)e^{-i\vec{Q}.\vec{r}}d\vec{r}\label{eq:fB(Q)}
\end{equation}

It can be shown that only the component of $\vec{f}_{B}\left(\vec{Q}\right)$
perpendicular to $\vec{Q}$ contributes to the scattering \cite{bruckel2009}:

\begin{equation}
\vec{f}_{B\bot}=\vec{q}\times\vec{f}_{B}\times\vec{q}=\vec{f}_{B}-\left(\vec{f}_{B}\cdot\vec{q}\right)\vec{q}\label{eq:fBperp(Q)}
\end{equation}

with $\vec{q}=\frac{\vec{Q}}{|Q|}$ being the unit vector along the
$\vec{Q}$ direction.

The total scattering cross section writes:

\begin{equation}
\frac{d\sigma}{d\Omega}\propto\left|f_{N}\left(\vec{Q}\right)+\vec{f}_{B\bot}\left(\vec{Q}\right)\cdot\vec{\sigma}\right|^{2}\label{eq:dsigma/dOmega}
\end{equation}

where we recall that $\vec{\sigma}$ is the neutron spin operator.

In the case of polarized neutrons with polarization analysis, it is
possible to measure 4 quantities corresponding to ``Non Spin Flip''
(NSF, (++) or (- -)) and ``Spin Flip'' (SF, (+-) or (-+)) scattering
as recently experimentally put into evidence \cite{honecker,Krycka2010}:

\begin{equation}
\begin{array}{l}
\begin{array}{c}
\frac{d\sigma^{++}}{d\Omega}=\left|f_{N}(\vec{Q})+f_{B_{\bot z'}}(\vec{Q})\right|^{2}\\
\frac{d\sigma^{--}}{d\Omega}=\left|f_{N}(\vec{Q})-f_{B_{\bot z'}}(\vec{Q})\right|^{2}\\
\frac{d\sigma^{+-}}{d\Omega}=\left|f_{B_{\bot x'}}(\vec{Q})-i\: f_{B_{\bot y'}}(\vec{Q})\right|^{2}\\
\frac{d\sigma^{-+}}{d\Omega}=\left|f_{B_{\bot x'}}(\vec{Q})+i\: f_{B_{\bot y'}}(\vec{Q})\right|^{2}
\end{array}\end{array}\label{eq:sigma PA}
\end{equation}

where $f_{B_{\bot x'}}(\vec{Q})$, $f_{B_{\bot y'}}(\vec{Q})$ and
$f_{B_{\bot z'}}(\vec{Q})$ refer to the component of $\vec{f}_{B_{\bot}}\left(\vec{Q}\right)$
along the $(Ox')$, $(Oy')$ and $(Oz')$ axis respectively, where
$(Oz')$ is the quantification axis of the neutron spin defined by
the applied magnetic field. If the induction distribution is even
around the $(Ox')$ and $(Oy')$ axes, the Fourier transforms $f_{B_{\bot x'}}(\vec{Q})$
and $f_{B_{\bot y'}}(\vec{Q})$ are real so that the two spin-flip
cross sections are equal. These formulae are derived from Eq. \ref{eq:dsigma/dOmega}
using the Pauli matrices \cite{Pauli}.

In the case of polarized neutrons without polarization analysis which
in practice is the situation encountered in most PSANS experiments,
the two scattering intensities which can be measured are: until now,
2 quantities can be measured:

\begin{equation}
\begin{array}{l}
\frac{d\sigma^{+}}{d\Omega}=\frac{d\sigma^{++}}{d\Omega}+\frac{d\sigma^{+-}}{d\Omega}=\\
\left|f_{N}(\vec{Q})+f_{B_{\bot z'}}(\vec{Q})\right|^{2}+\left|f_{B_{\bot x'}}(\vec{Q})+i\: f_{B_{\bot y'}}(\vec{Q})\right|^{2}\\
\\
\frac{d\sigma^{-}}{d\Omega}=\frac{d\sigma^{--}}{d\Omega}+\frac{d\sigma^{-+}}{d\Omega}=\\
\left|f_{N}(\vec{Q})-f_{B_{\bot z'}}(\vec{Q})\right|^{2}+\left|f_{B_{\bot x'}}(\vec{Q})-i\: f_{B_{\bot y'}}(\vec{Q})\right|^{2}
\end{array}\label{eq:dsigma pol neutrons}
\end{equation}

For non polarized neutron scattering: 
\begin{equation}
\begin{array}{l}
\frac{d\sigma}{d\Omega}=\frac{d\sigma^{+}}{d\Omega}+\frac{d\sigma^{-}}{d\Omega}=\left|f_{N}\left(\vec{Q}\right)\right|^{2}+\left|\vec{f}_{B\bot}(\vec{Q})\right|^{2}\end{array}
\end{equation}

\subsection{Practical calculation of the nuclear and magnetic form factors}

We shall now discuss how to calculate in practice the quantity $\vec{f}_{B\bot}(\vec{Q})$.
A first approach consists in assuming an a priori knowledge of the
magnetization distribution in the particle (for example, $\vec{M}$
can be assumed to be homogeneous). Under this assumption, the induction
field distribution $\overrightarrow{B}\left(\vec{r}\right)$ can be
calculated using standard electromagnetic softwares. In the case of
objects with an axis of revolution, we suggest the use of the \emph{Femm}
package \cite{FEMM}. The package provides various tools (scripting
langage \emph{LUA}, \emph{Octavefemm} interface to \emph{Octave} or
\emph{Matlab}, \emph{Mathfemm} interface to \emph{Mathematica}) which
allows extracting the induction field distribution $\overrightarrow{B}\left(\vec{r}\right)$
from the electromagnetic calculation.\\

A more general approach, with no assumptions on the magnetization
distribution in the nanoparticle calculates the magnetization distribution
in the particle under a given applied magnetic field. Several micromagnetic
packages have become available for non expert users during the last
few years (\emph{OOMMF} \cite{oommf}, \emph{MagPar} \cite{magpar},
\emph{Nmag} \cite{nmag}). We have used the \emph{Nmag} package since
it is based on finite elements and is thus especially well suited
to the types of particles encountered in SANS scattering (spheres,
cylinders, disks). The procedure consists in defining an object geometry
and discretization using the \emph{Netgen} mesher \cite{netgen} (Figure
\ref{Fig3:models}). The 3D magnetic moments distribution $\overrightarrow{M}\left(\vec{r}\right)$
in the object can then be calculated using the \emph{Nmag} package
(Figure \ref{Fig3:models}e).

In order to perform a numerical calculation of the magnetic form factors
of such magnetic objects, a volume of space $V$ containing the objects
is defined. Its dimensions $L_{x}\times L_{y}\times L_{z}$ are such
that this volume is significantly large than the nanoobjects (3-10
times). From the magnetization distribution $\overrightarrow{M}(\vec{r})$
in the nano-object, the induction field $\overrightarrow{B}\left(\vec{r}\right)$
can be calculated in the whole volume $V$. In practice, this induction
field $\overrightarrow{B}\left(\vec{r}\right)$ is calculated on a
regular grid and mapped on a 3D matrix of size $(n_{x},n_{y},n_{z})$.
Python scripts performing this task are available on the LLB Website
\cite{Scripts}. This induction distribution is then exported into
\emph{octave} or \emph{matlab} in order to perform numerical calculations
of the magnetic form factor $\vec{f}_{B}(\vec{Q})$ using equations
(\ref{eq:fB(Q)}) and (\ref{eq:fBperp(Q)}). The corresponding scripts
are freely available at \cite{Scripts}. One obtains a set of three
3D matrices $\left\{ f_{Bx}(\vec{Q}),\, f_{By}(\vec{Q}),\, f_{Bz}(\vec{Q})\right\} $
describing the 3 components of the magnetic form factor in the reciprocal
space. The reciprocal space is mapped with a sampling given by $\Delta Q=\frac{2\pi}{L_{i}}$
and the useful accessible $Q$-range goes from $Q_{min}=\frac{2\pi}{L_{i}}$
to about $Q_{i}^{max}\sim0.5\times\frac{2\pi}{L_{i}}n_{i}$ (e.g.
for $L_{i}=100nm$ and $n_{i}=100$, $0.06<Q<3nm^{-1}$ ) \\

The Fourier transforms of the induction field are performed using
a FFT algorithm with the condition that the dimensions $(n_{x},n_{y},n_{z})$
of the matrices are powers of 2. Since we are dealing with 3D matrices,
the memory requirements diverge rather quickly as the matrices sizes
increase. A matrix size of $128\times128\times128$ provides both
a reasonable memory footprint (32 MBytes / matrix) and fast calculations
times ($50ms$ per FFT on a standard desktop computer). Such a sampling
rate provides accurate Fourier transforms in the reciprocal space
over 2 orders of magnitude (see Figure 5 for example). We found that
a volume of space $V$ about 6 times larger (i.e. $L_{x}=6D$ for
a sphere of diameter $D$) than the studied object provided good results
in the $Q$ range of interest.\\
\begin{figure*}
\includegraphics[bb=60bp 360bp 680bp 550bp,clip,width=15cm]{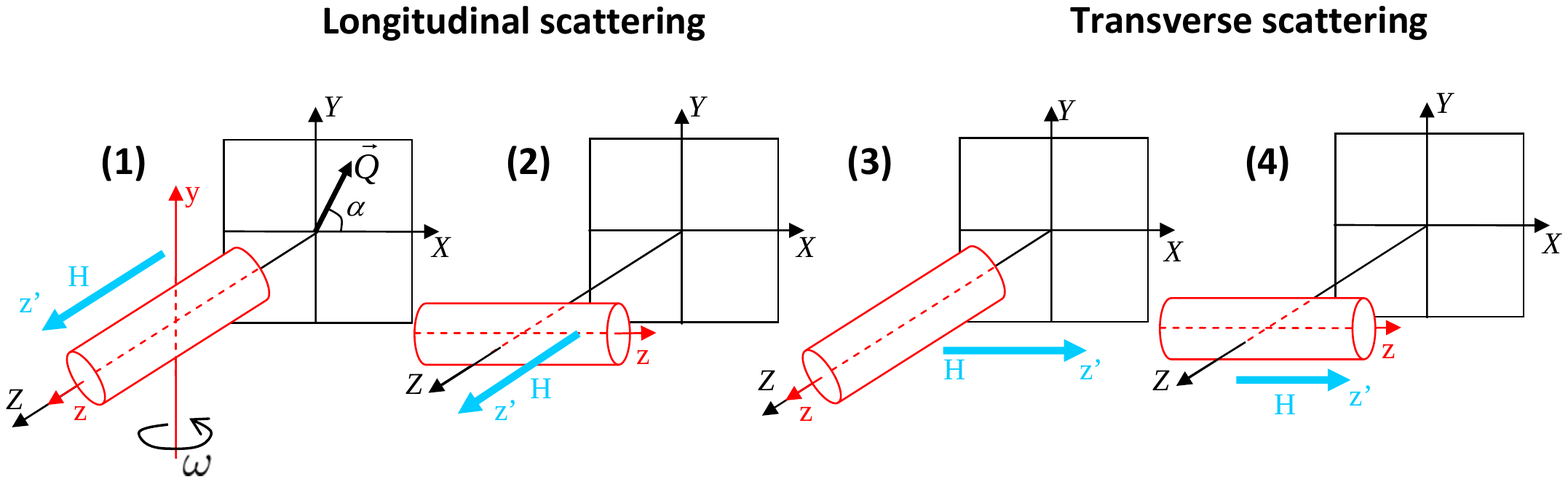}

\caption{The different scattering that will be considered. The neutron beam
is incident along the $(OZ)$ direction. $(XY)$ defines the scattering
plane. The longitudinal\emph{ }(1-2) and transverse (3-4) scattering
geometries correspond to a magnetic field applied along $(OZ)$ and
$(OX)$ respectively. The scattering object can be oriented with its
easy axis parallel (1-4) or perpendicular (2-3) to the applied field.}

\label{Flo:scatt geom} 
\end{figure*}

We now define the scattering geometries that will be considered in
the actual numerical calculations presented in the next section. We
first underline that in the general case of polarized neutron scattering
on anisotropic particles, it is necessary to define 3 axis systems.
The first one $(x,y,z)$ is attached to the studied nano-objects.
The second one $(X,Y,Z)$ describes the spectrometer geometry. The
third one $(x',y',z')$ describes the orientation of the quantification
axis (defined by the applied magnetic field along $z'$). In order
to describe a scattering experiment, it is necessary to define the
relative orientations of these 3 sets of axis.

In order to simplify the presentation and to comply with experimental
conditions, we consider a scattering plane $\left(Q_{X}=Q\cos\alpha,\: Q_{Y}=Q\sin\alpha\right)$
corresponding to the 2D detector plane of a SANS spectrometer $(XOY)$.
For each scattering direction $\alpha$ in the scattering plane, the
perpendicular components of the magnetic form factor $\vec{f}_{\bot}\left(\theta,r\right)$
can be calculated using equation (\ref{eq:fBperp(Q)}):

\begin{equation}
\left(\begin{array}{c}
f_{B_{\perp X}}\\
f_{B_{\perp Y}}\\
f_{B_{\perp Z}}
\end{array}\right)=\left(\begin{array}{c}
f_{B_{X}}-(f_{B_{X}}\cos\alpha+f_{B_{Y}}\sin\alpha)\cos\alpha\\
f_{B_{Y}}-(f_{B_{X}}\cos\alpha+f_{B_{Y}}\sin\alpha)\sin\alpha\\
f_{B_{Z}}
\end{array}\right)\label{eq:fbperp components}
\end{equation}

Note that the above formula is only valid in the $(Q_{X},\: Q_{Y})$
scattering plane. In the practical procedure, the magnetic induction
distribution is calculated in the $(xyz)$ reference frame where $(Oz)$
is the revolution axis of the object. The examples proposed in \cite{Scripts}
follow this convention. We define $\omega$ as the rotation angle
of the object around the $(OY)$ axis. With these conventions, for
$\omega=0$, a cylinder is aligned along the $(OZ)$ axis. For $\omega=90\lyxmathsym{\textdegree}$,
a cylinder is aligned along the $(OX)$ axis. The distribution $\overrightarrow{B}\left(\vec{r}\right)_{XYZ}$
in the spectrometer axis is obtained by a rotation $\omega$ :

\begin{equation}
\left(\begin{array}{c}
B_{\perp X}\\
B_{\perp Y}\\
B_{\perp Z}
\end{array}\right)=\left(\begin{array}{ccc}
cos\omega & 0 & sin\omega\\
0 & 1 & 0\\
-sin\omega & 0 & cos\omega
\end{array}\right)\left(\begin{array}{c}
B_{\perp x}\\
B_{\perp x}\\
B_{\perp z}
\end{array}\right)\label{eq:rotation}
\end{equation}

For the sake of illustration, we shall not include both the nuclear
scattering and the magnetic scattering in the calculations. Even though
this is not realistic in a general scattering experiment where the
nuclear contribution is usually large, such situations can be encountered
in SANS experiments: the nuclear contrast of cobalt particles can
easily be matched by using an appropriate deuterated solvent \cite{Wiedenmann2003,Wiedenmann2004};
ferromagnetic particles in a paramagnetic matrix also allow to extinguish
the nuclear contrast \cite{michels2006}. Ignoring the nuclear scattering
effects will allow us to focus on the magnetic effects which are the
scope of this communication.

\begin{figure*}
\includegraphics[bb=50bp 310bp 750bp 560bp,clip,width=15cm]{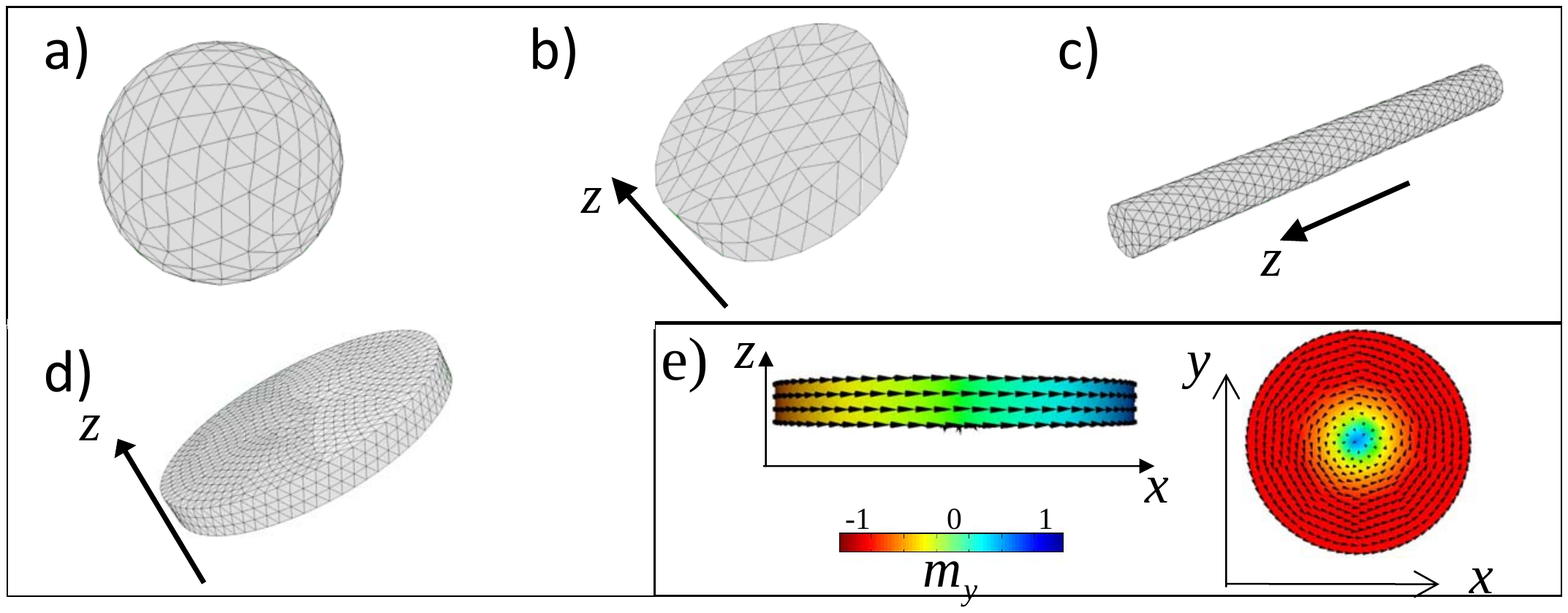}

\caption{Typical meshes obtained from the \emph{Netgen} mesher for the different
nano-objects: (a) a sphere of diameter $D=10$ nm, (b) a flat disk
of diameter $D=20$ nm and height $h=2.5$ nm, (c) a cylinder of diameter
$D=10$ nm and length $L=50$ nm, (d) a disk of diameter $D=50$ nm
and height $h=5$ nm. Note that the distance between two nodes is
around 1 nm. (e) Maps of the magnetic moments distribution at remanence
for the disk defined in (d). The colors encode the $y$-component
of the magnetization ($m_{y}=M_{y}/M$) in the profile view (left)
and $z$-component of the magnetization ($m_{z}=M_{z}/M$) viewed
from the top (right).}

\label{Fig3:models} 
\end{figure*}

In the following, we shall different relative orientations of the
applied magnetic field direction with respect to the scattering plane: 
\begin{enumerate}
\item \emph{Longitudinal} scattering $(H\parallel OZ)$ . Equation (\ref{eq:sigma PA})
becomes: 
\begin{equation}
\begin{array}{l}
\frac{d\sigma^{++}}{d\Omega}=\frac{d\sigma^{--}}{d\Omega}=\left|f_{B_{\bot Z}}(\vec{Q})\right|^{2}\\
\frac{d\sigma^{+-}}{d\Omega}=\frac{d\sigma^{-+}}{d\Omega}=\left|f_{B_{\bot X}}(\vec{Q})\mp i\: f_{B_{\bot Y}}(\vec{Q})\right|^{2}
\end{array}\label{eq:Cross Sections Long}
\end{equation}

\item \emph{Transverse} scattering $(H\parallel OX)$ . Equation (\ref{eq:sigma PA})
becomes: 
\begin{equation}
\begin{array}{l}
\frac{d\sigma^{++}}{d\Omega}=\frac{d\sigma^{--}}{d\Omega}=\left|f_{B_{\bot X}}(\vec{Q})\right|^{2}\\
\frac{d\sigma^{+-}}{d\Omega}=\frac{d\sigma^{-+}}{d\Omega}=\left|f_{B_{\bot Y}}(\vec{Q})\mp i\: f_{B_{\bot Z}}(\vec{Q})\right|^{2}
\end{array}\label{eq:Cross Sections Trans}
\end{equation}

\end{enumerate}
Note that even though the above formulae are very similar, the key
parameter is the distribution $f_{B_{\bot}}(\vec{Q})$ with respect
to the scattering plane. The scattering geometries (1)-(4) presented
on Figure \ref{Flo:scatt geom} will lead to qualitatively different
results which are discussed in the next section.

\section{Model calculations}

In this section we will present the calculation of the magnetic form
factors of different types of nano-objects: (a) a sphere of diameter
$D=10$ nm, (b) a flat disk of diameter $D=10$ nm and height $h=2.5$
nm and (c) a cylinder of diameter $D=10$ nm and length $L=50$ nm.
These objects are presented in Figure \ref{Fig3:models}. The magnetic
parameters used are the following: magnetization saturation of cobalt
$M=1.4\times10^{6}$ A.m$^{-1}$($\equiv$1.76 T); exchange stiffness
$A=1.2\times10^{11}$ J.m$^{-1}$. Moreover, a uniaxial magnetocrystalline
anisotropy $K=7\times10^{5}$ J.m$^{-3}$ along $z$ was considered
in the case of the flat disk (Figure \ref{Fig3:models}b) in order
to force the magnetic moments to be normal to the disk surface. In
a first step, the remanent configuration was determined for all the
objects. It reveals a rather homogeneous distribution of the magnetic
moments along the $Oz$ direction for objects (a), (b) and (c). This
is due to the fact that these objects are to the first order approximations
of elliptical objects in which the demagnetizing field in homogeneous.
For simplicity, only nano-objects with a quasi uniform magnetic moments
distribution have been considered. However, more complex distributions
can be considered as it is highlight for a disk of diameter $D=50$
nm and height $h=5$ nm within which a vortex state appears (see Figure
\ref{Fig3:models}e).

We perform numerical calculation of the 3D nuclear form factors so
as to be able to compare the results with analytical formulae and
validate the numerical procedure. As a next step, we perform calculations
of the magnetic form factors where we illustrate the contributions
of the demagnetizing and the dipolar fields.

\subsection{Validation of the numerical procedure: application to the calculation
of nuclear form factors}

Although this paper is devoted to the magnetic form factors of nano-objects,
we have also numerically calculated the 3D nuclear form factor of
these objects in order to check the applicability of our numerical
algorithm. When calculating the 3D nuclear form factor, the volume
of space $V$ containing the object is mapped onto a 3D matrix $G_{x,y,z}$
describing the object geometry in such a way that $G_{x,y,z}=1$ for
$(x,y,z)$ inside the object and $G_{x,y,z}=0$ for $(x,y,z)$ outside
the object.

\begin{figure*}
\includegraphics[bb=60bp 200bp 790bp 560bp,clip,width=15cm]{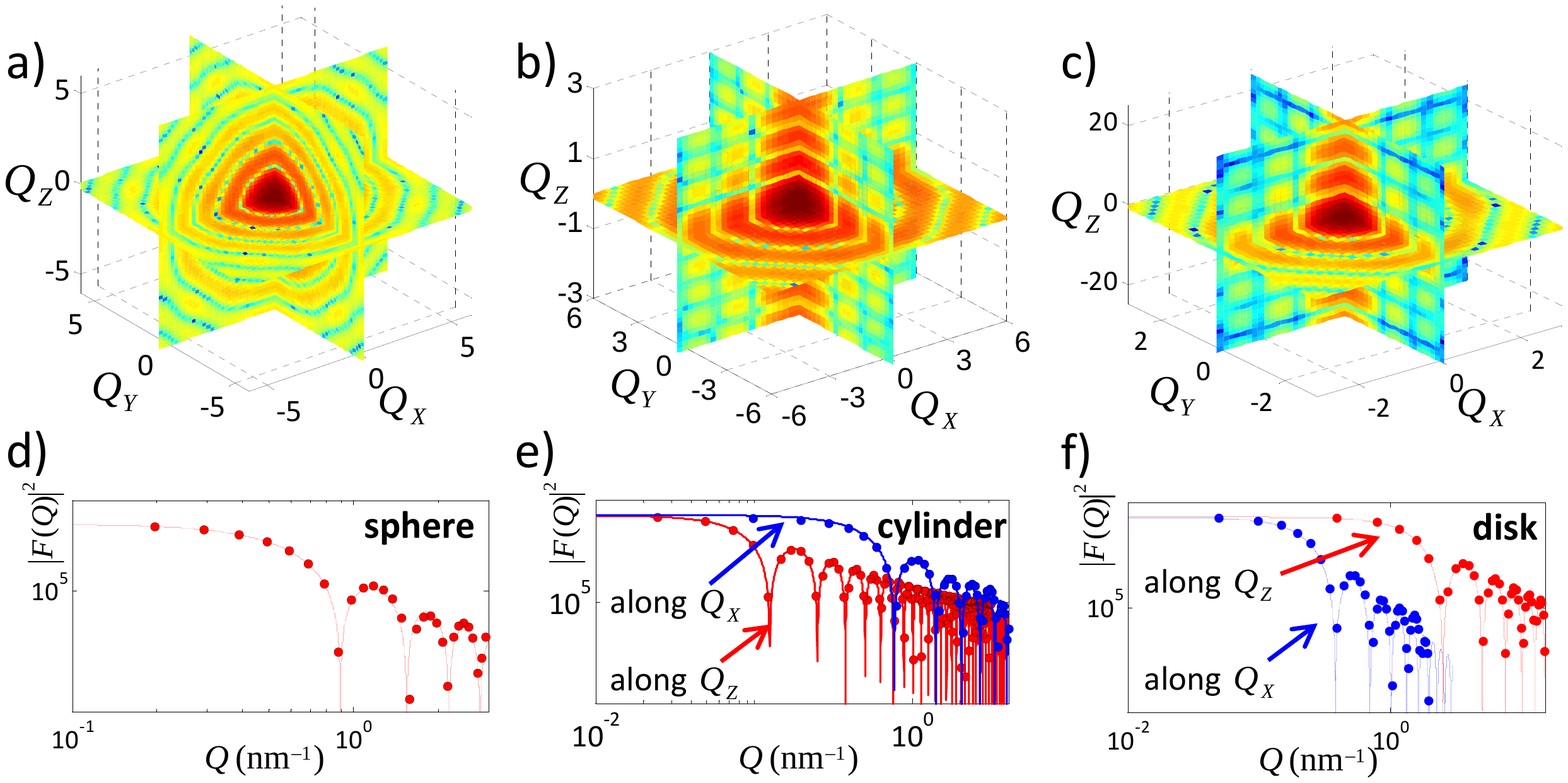}

\caption{Numerically calculated form factors (circles) and comparison with
analytical formulas (continuous lines). a), b) and c) correspond to
3D form factors calculated for the sphere $D=10$, the cylinder ($D=10$
nm, $L=50$ nm), the flat disk ($D=10$ nm, $h=2.5$ nm) and 1D cuts
of the form factor in the reciprocal space in specific directions.}

\label{Fig4:FN} 
\end{figure*}

Figures \ref{Fig4:FN}a-b-c present the 3D nuclear form factors calculated
for the sphere ($D=10$ nm), the flat disk ($D=10$ nm, $h=2.5$nm)
and the cylinder ($D=10$ nm, $L=50$ nm) defined in Figure \ref{Fig3:models}.
The scattering is isotropic in the sphere case and anisotropic in
the other cases. The nuclear form factors of these simple objects
are well known and can be analytically calculated \cite{glatter}.
We can thus compare our numerical results with the analytical formulae.
Figures \ref{Fig4:FN}d-e-f compare the numerical and analytical calculations
along the main directions in the reciprocal space (in nm$^{-1}$)
for the different objects. The agreement is satisfactory for all the
objects which proves the applicability of our numerical calculations.

These results show that it is possible to calculate rather accurately,
in a single shot, the form factor in a $Q$-range extending over 2
decades with scattering intensities extending over 3 decades. If one
would be interested in a different $Q$-range, it is necessary to
resize the \emph{$V$} box around the object. As $Q_{min}=\frac{2\pi}{L_{i}}$,
in order to probe smaller Q values, the size $L_{i}$ of the box should
be increased As $Q_{i}^{max}\sim0.5\times\frac{2\pi}{L_{i}}n_{i}$,
in order to probe larger Q values, either the size $L_{i}$ of the
box should be decreased of the number of mapping points $n_{i}$ should
be increased. This is similar to a real SANS experiment where the
measuring conditions are changed in order to cover a wider $Q$-range.

In the following we shall however restrict ourselves to a $V$ box
$6\times6\times6$ times the size of the object, which provides an
appropriate sampling in the $Q$ space for our demonstrations.

\begin{figure*}
\includegraphics[bb=60bp 215bp 710bp 550bp,clip,width=15cm]{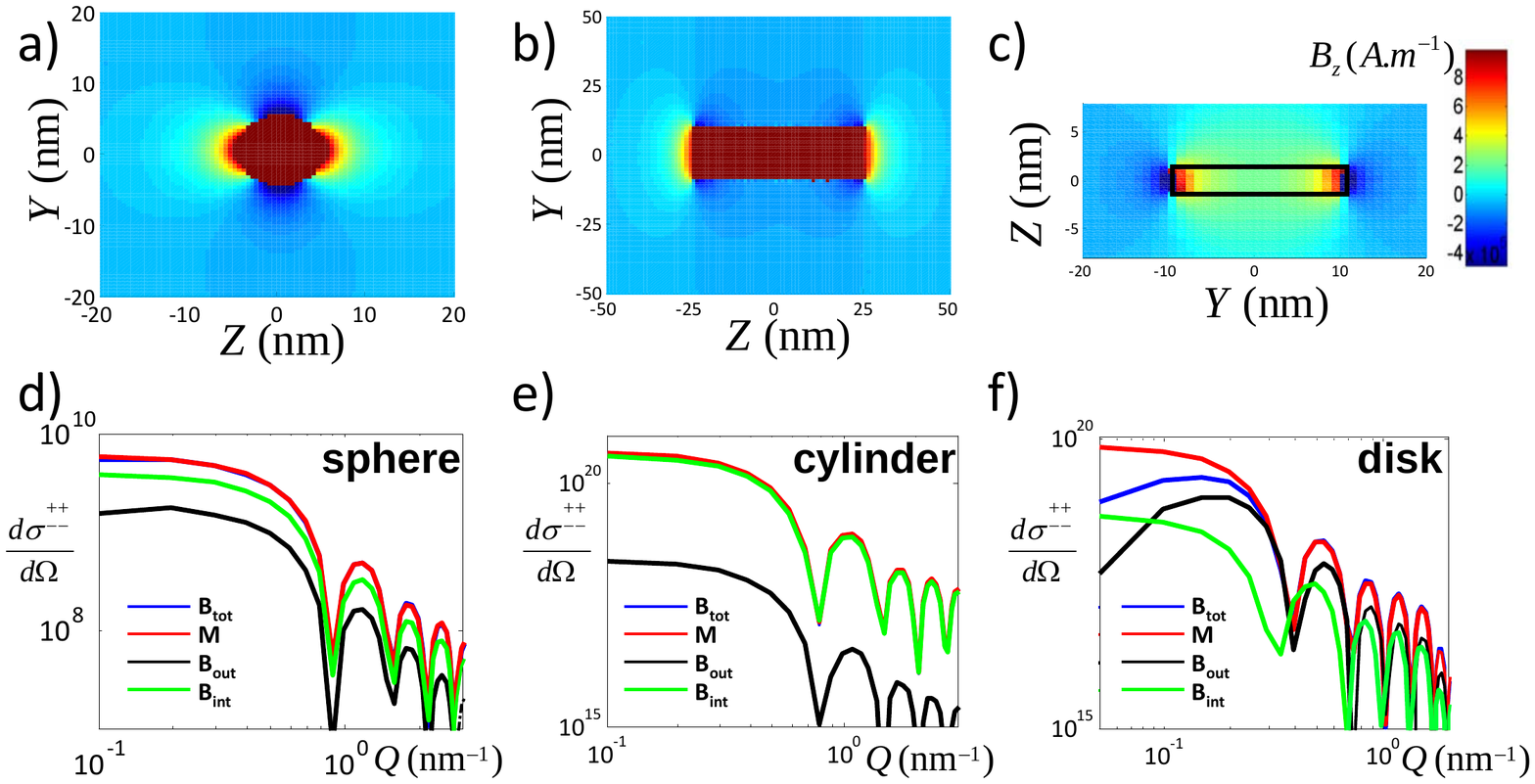}

\caption{Longitudinal scattering on 3 types of objects (sphere, cylinder and
disk) magnetized along the applied field. a), b), and c) Induction
distribution in ($YZ$) plane for a sphere, a cylinder and a disk
magnetized along the $Z$ direction. d), e) and f) Scattering cross
sections of different magnetic contributions: the magnetization only
($\vec{M}(\vec{r)}$), the internal magnetic induction ($\vec{B}_{int.}(\vec{r)}$),
the dipolar field created by the magnetic moments outside the object
($\vec{B}_{ext.}(\vec{r)}$) and the whole magnetic induction ($\vec{B}{}_{tot.}(\vec{r)}=\vec{B}_{int.}(\vec{r)}+\vec{B}_{ext.}(\vec{r)}$).}
\end{figure*}

\subsection{Case of magnetic objects}

The same calculation procedure was applied to the case of magnetic
objects. Figure 5 presents the calculation of the magnetic form factors
in the case of the 3 canonical geometries (sphere, disk and cylinder).
A first reference calculation was performed by calculating the form
factor of an homogeneously magnetized sample and for which only the
magnetization inside the sample was considered (blue lines on Figure
5). A second calculation was performed where the form factor of the
induction map was also included. The induction map is plotted on Figure
5 a-c. We considered the induction in the sample given by $\vec{B}_{int.}(\vec{r)}=\mu_{0}\left(\vec{M}(\vec{r)}+\vec{H}_{d}(\vec{r)}\right)$
where $\vec{H}_{d}(\vec{r)}$ is the demagnetizing fields and the
stray fields outside the sample ($\vec{B}_{ext.}(\vec{r)}$). The
contribution of these inductions distributions was calculated separately
(black and green curves). The sum of these contributions $\vec{B}_{tot.}(\vec{r)}$
overlaps with the form factor simply obtained from the magnetization
distribution (see figure 5d-e-f). The deviation observed at low $Q$
are numerical artifacts due to the fact that the stray fields expand
beyond the calculation box. This numerically demonstrates that the
magnetic form factor can be obtained by considering the magnetization
distribution without taking into account the demagnetization fields
and the stray fields outside the sample.

\section{Conclusion}

We have developed numerical tools \cite{Scripts} which permits the
calculations of the structural and magnetic form factor of nano-objects
with complex geometrical shapes. Such calculations were until recently
limited due to the extensive memory usage of 3 dimensional FFT calculations.
As Polarized SANS spectrometers are becoming widely available the
problem of quantitatively processing the data obtained on complex
magnetic nano-systems will arise. This first step will allow experimentalists
to compare their measurements with realistic models of the magnetization
in their nano-systems. In the future, with the widespread availability
of systems with tens of gigabytes of memory, one may consider applying
similar methods for the numerical calculation of complex structure
factors.
\begin{acknowledgments}
The authors gratefully acknowledge the Agence Nationale de la Recherche
for their financial support (project P-Nano MAGAFIL). We thank the
\emph{Nmag} developers for their advices. This research project has
also been supported by the European Commission under the 7th Framework
Programme through the 'Research Infrastructures' action of the 'Capacities'
Programme, NMI3-II Grant number 283883. \end{acknowledgments}


\begin{thebibliography}{10}
\bibitem{wade2005}T. L. Wade, J. E. Wegrowe, European Physical Journal
- Applied Physics, \textbf{29}, 3 (2005)

\bibitem{soulantica2009}K. Soulantica, F. Wetz, J. Maynadie et al.,
Appl. Phys. Lett. \textbf{95}, 152504 (2009).

\bibitem{viau2008} G. Viau, C. Garcia, T. Maurer et al. Phys. Stat.
Sol. A - Applications and Materials Science \textbf{206} 663-666 (2009).

\bibitem{soumare2008}Y. Soumare, J. -Y. Piquemal, T. Maurer et al.
J. of Materials Chemistry \textbf{18}, 5696-5702 (2008).

\bibitem{soumare2009}Y. Soumare, C. Garcia, T. Maurer et al., Advanced
Functional Materials \textbf{19,} 1971 (2009)

\bibitem{michels_rev}A. Michels and J. Weissmï¿œller, Rep. Prog.
Phys., \textbf{71} 066501 (2008)

\bibitem{Wiedenmann2003} A. Hoell, M. Kammel, A. Heinemann and A.
Wiedenmann, J. Appl. Cryst. \textbf{36}, 558 (2003)

\bibitem{Wiedenmann2004} K. Butter, A. Hoell, A. Wiedenmann, A. V.
Petukhov and G.-J. Vroege, J. Appl. Cryst. \textbf{37}, 847 (2004)

\bibitem{wiedenmann2006}A. Wiedenmann , M. Kammel , A. Heinemann
and U. Keiderling, Journal of Physics: Condensed Matter \textbf{18,}
S2713 (2006)

\bibitem{michels2005}A. Michels, C. Vecchini, O. Moze, K. Suzuki,
J. M. Cadogan, P. K. Pranzas and J. Weissmœller, Europhys. Lett.,
\textbf{72} (2), 249 (2005)

\bibitem{michels2006} A. Michels, C. Vecchini and O. Moze, K. Suzuki,
P. K. Pranzas, J. Kohlbrecher, J. Weissmœller, Physical Review B,
\textbf{74,} 134407 (2006)

\bibitem{bruckel2009}T. Bruckel, G. Heger, D. Richter and R. Zorn,
Neutron Scattering - Lectures of the Laboratory Course held at the
FZ Jœlich, Vol. 28 (2009) p. 3-22. ISSN 1433-5506 ISBN 3-89336-395-5.
http://hdl.handle.net/2128/418

\bibitem{honecker}D. Honecker, A. Ferdinand, F. Dobrich, C. D. Dewhurst,
A. Wiedenmann, C. Gomez-Polo, K. Suzuki and A. Michels, Eur. Phys.
J. B (2010), DOI: 10.1140/epjb/e2010-00191-5

\bibitem{Krycka2010}K. L. Krycka, R. A. Booth, C. R. Hogg, Y. Ijiri,
J. A. Borchers, W. C. Chen, S. M.Watson, M. Laver, T. R. Gentile,
L. R. Dedon, S. Harris, J. J. Rhyne and S. A. Majetich, Phys. Rev.
Lett., \textbf{104}, 207203 (2010)

\bibitem{Pauli}$\vec{\sigma}=\left(\sigma_{x},\sigma_{y},\sigma_{z}\right)$;
Pauli operator rules: $\sigma_{x}\left|+\right\rangle =\left|-\right\rangle $,
$\sigma_{x}\left|-\right\rangle =\left|+\right\rangle $, $\sigma_{y}\left|+\right\rangle =-i\left|-\right\rangle $,
$\sigma_{y}\left|-\right\rangle =i\left|+\right\rangle $, $\sigma_{z}\left|+\right\rangle =\left|+\right\rangle $,
$\sigma_{z}\left|-\right\rangle =\left|-\right\rangle $

\bibitem{FEMM}D. C. Meeker, Finite Element Method Magnetics. The
FEMM package is freely available at http://www.femm.info.

\bibitem{oommf} M. J. Donahue, D. G. Porter, R. D. McMichael and
J. Eicke, Public code OOMMF URL: http://math.nist.gov/oommf

\bibitem{magpar}MagPar: http://www.magpar.net/

\bibitem{nmag}T. Fischbacher, M. Franchin, G. Bordignon and H. Fangohr,
IEEE Trans. Mag. \textbf{43,} 2896 (2007). URL : http://nmag.soton.ac.uk/nmag/

\bibitem{netgen}NETGEN, automatic mesh generator, URL: http://www.hpfem.jku.at/netgen/

\bibitem{Scripts}http://www-llb.cea.fr/MagneticFormFactors/

\bibitem{glatter} O. Glatter and O. Kratky, Small Angle X-Ray Scattering,
Academic Press, 1982.\end{thebibliography}
\end{document}